\documentclass[a4paper,11pt,twoside]{article}
\usepackage{amssymb,amsmath,amsthm}
\usepackage[ps,dvips,curve,knot,frame,matrix]{xy}

\renewcommand{\k}{\Bbbk}

\newcommand{\Z}{\mathbb{Z}}

\newcommand{\R}{\mathbb{R}}
\newcommand{\C}{\mathbb{C}}
\DeclareMathOperator{\id}{id}
\newcommand{\tens}{\otimes}
\newcommand{\defeq}{:=}
\DeclareMathOperator{\cop}{\Delta}
\DeclareMathOperator{\cou}{\epsilon}
\DeclareMathOperator{\antip}{\mathrm{S}}
\newcommand{\act}{\triangleright}
\newcommand{\Ru}{\mathcal{R}}
\renewcommand{\i}[1]{{}_{\scriptscriptstyle(#1)}}
\newcommand{\iu}[1]{{}_{\scriptscriptstyle(\underline #1)}}
\newcommand{\ib}[1]{{}_{\scriptscriptstyle[#1]}}
\newcommand{\ibu}[1]{{}_{\scriptscriptstyle[\underline #1]}}
\newcommand{\catmod}[1]{{}^{#1}\!{\mathcal{M}}}
\newcommand{\nateq}{\sigma}
\newcommand{\funceq}{\mathcal{F}}
\newcommand{\falg}{\mathcal{C}}

\renewcommand{\O}{\mathcal{O}}
\newcommand{\xd}{\mathrm{d}}
\newcommand{\Rd}{\R^d}
\newcommand{\Td}{\mathbb{T}^d}
\newcommand{\UR}{\mathsf{U}(\Rd)}
\newcommand{\URt}{\mathsf{U}_\theta(\Rd)}
\newcommand{\URc}{\widetilde{\mathsf{U}}(\Rd)}
\newcommand{\URct}{\widetilde{\mathsf{U}}_\theta(\Rd)}
\newcommand{\CR}{\mathcal{C}(\Rd)}
\newcommand{\CRt}{\mathcal{C}_\theta(\Rd)}
\newcommand{\CRmt}{\mathcal{C}_{-\theta}(\Rd)}
\newcommand{\CRc}{\widetilde{\mathcal{C}}(\Rd)}
\newcommand{\CRct}{\widetilde{\mathcal{C}}_\theta(\Rd)}

\newcommand{\pdiff}[1]{\frac{\partial}{\partial #1}}
\newcommand{\CT}{\mathcal{C}(\Td)}
\newcommand{\CTct}{\widetilde{\mathcal{C}}_\theta(\Td)}
\newcommand{\chit}{\chi_\theta}
\newcommand{\chimt}{\chi_{-\theta}}
\newcommand{\Rut}{\mathcal{R}_\theta}
\newcommand{\psit}{\psi_\theta}
\newcommand{\psimt}{\psi_{-\theta}}

\theoremstyle{plain}
\newtheorem{prop}{Proposition}[section]
\newtheorem{lem}[prop]{Lemma}
\newtheorem{thm}[prop]{Theorem}

\newtheorem{ex}[prop]{Example}

\newcommand{\sxy}[1]{{\begin{xy} #1 \end{xy}}}
\everyxy={0;<0.6mm,0mm>:<0mm,0.75mm>::0;0,}

\begin{document}
\title{\textbf{Untwisting Noncommutative $\Rd$ and the Equivalence of
Quantum Field Theories}}
\author{Robert Oeckl\footnote{email: r.oeckl@damtp.cam.ac.uk}\\ \\
Department of Applied Mathematics and Mathematical Physics,\\
University of Cambridge,
Cambridge CB3 0WA, UK}
\date{DAMTP-2000-26\\
2 March 2000, 11 March 2000 (v2)}

\maketitle

\vspace{\stretch{1}}

\begin{abstract}
We show that there is a duality exchanging noncommutativity and
non-trivial statistics for quantum field theory on $\Rd$.
Employing methods of quantum groups, we observe that ordinary
and noncommutative $\Rd$ are related by twisting.
We extend the twist to an equivalence for quantum field theory
using the framework of braided quantum field theory.
The twist exchanges both commutativity with noncommutativity and
ordinary with non-trivial statistics.
The same holds for the noncommutative torus.
\end{abstract}

\vspace{\stretch{1}}

\clearpage
\section*{Introduction}

Quantum field theory with
noncommuting coordinates was proposed a long time ago with the hope to
regularise divergencies
\cite{Sny:quantized}. A more ambitious
motivation comes from the possible role of noncommutative geometry
in the ongoing struggle to unite gravity with
quantum field theory.
Needless to say, the issue of even defining a quantum field theory on
a noncommutative space is highly nontrivial.
However, a generalisation of quantum field theory to
noncommutative spaces with symmetries has recently been proposed
\cite{Oe:bQFT}.

In this paper we consider
coordinate commutation relations of the form
\begin{equation*}
 [x^\mu,x^\nu]=i\,\theta^{\mu\nu}
\end{equation*}
in $d$ dimensions,
where $\theta$ is a real-valued antisymmetric matrix.
This can also be viewed as equipping the algebra of functions on
$\Rd$ with a deformation quantised multiplication known
as a Moyal $\star$-product \cite{Moy:quantstat}.
We refer to this space-time algebra as
``noncommutative $\Rd$''.
We shall also consider the toroidal compactification 
known as the ``noncommutative torus''. It served as an early example
of a noncommutative geometry for Connes \cite{Con:cstargeom}.

Commutation relations of the type considered
here were proposed 
by Doplicher, Fredenhagen and Roberts 
based on an analysis of the constraints posed by general relativity
and Heisenberg's uncertainty principle
\cite{DoFrRo:qsfields}.
(For other approaches at noncommutative space-times see
\cite{Maj:hopfplanck,Mad:fuzzyphys,PoWo:qlorentz,LNR:kpoincare}.)
They also initiated the study of quantum field
theory on this kind of space-time.
(For an alternative approach to quantum field theory with
generalised uncertainty relations see \cite{Kem:uncertain}).
Basic results for Feynman diagrams relating the noncommutative and
the commutative setting were obtained by Filk \cite{Fil:qspace}.
With the emergence of the noncommutative torus in string theory
\cite{CoDoSc:nctori},
quantum field theory on such a space has received
a much wider interest, see \cite{SeWi:stringncg} and references
therein.
Recently, the perturbation theory has been of particular interest with
the investigation of divergencies and
renormalisability, see e.g.\ \cite{CR:renormnc,MVS:ncpert}.

Our treatment makes essential use of the theory of quantum groups and
braided spaces (see e.g.\ \cite{Ma:book}).
The starting point is the observation by Watts \cite{Wat:ncstringhopf}
that ordinary and noncommutative $\Rd$ are related by a certain
2-cocycle. This cocycle is
associated with the translation group (which we also denote by $\Rd$)
and induces a ``twist''. While the twist
turns $\Rd$ into itself as a group, it turns $\Rd$ into noncommutative
$\Rd$ as a representation.
Importantly, the concepts of cocycle and twist used here are
dual to those of ordinary group cohomology and arise only from the
quantum group point of view.
Employing the framework of braided quantum field theory \cite{Oe:bQFT}
enables us
to describe quantum field theory on both commutative and
noncommutative $\Rd$ in a purely algebraic language. This allows the
extension of the twist relating the two spaces to an equivalence
between the quantum field theories living on them. Underlying is
an equivalence of categories of representations.
However, the noncommutative $\Rd$ in this context carries a
non-trivial statistics.

The noncommutative $\Rd$ with ordinary
statistics (which is the space considered in the literature)
on the other hand is related by the same twist to commutative $\Rd$
with non-trivial statistics. Here as well, we obtain an equivalence of
quantum field theories on the two spaces.
In this case it is really a duality exchanging noncommutativity and
non-trivial statistics. In terms of
perturbation theory, the duality exchanges
a setting where vertices are noncommutative with a setting where
vertices are commutative, but crossings carry an extra Feynman rule.
As a byproduct, Filk's results are an immediate consequence.
Finally, we investigate further space-time symmetries and gauge
symmetry.
We find that while they are preserved by the twist (as quantum group
symmetries) they are broken by removing 
the non-trivial statistics from noncommutative $\Rd$.
Although the discussion is in terms of $\Rd$ for convenience, it
applies identically to the torus (except for the extra space-time
symmetries).

Our equivalence result also suggests that a noncommutativity of the kind
considered here really is to ``weak'' to be able to regularise a
quantum field theory. What one needs for that purpose is a
``stronger'' noncommutativity in the form of a strict braiding (with a
double-exchange not being the identity).
This is for example provided by $q$-deformations of Lie groups.
That quantum field theory can indeed be regularised in this way was
demonstrated in \cite{Oe:bQFT}.

Note that the concept of twisting has been used to relate
quantum space-times \cite{Ma:qwick} and quantum mechanical models
\cite{MaOe:twist} before. Also, 2-cocycles of ordinary group
cohomology have been used to obtain noncommutative spaces
in the context of matrix theory \cite{HoWu:ncmat}.

The article starts in Section~\ref{sec:twists} with the mathematical
basis concerning twisting, twist equivalence, and the relation with
deformation quantisation. This is mainly to equip the reader with the
structures, formulas, and statements required later and contains only
minimal new results.
Section~\ref{sec:Rd} looks at noncommutative $\Rd$ from the quantum
group point of view and establishes the equivalence with ordinary
$\Rd$ via twisting. The torus is treated as a special case.
The main part of the article is section~\ref{sec:qft}, where quantum field
theory on noncommutative $\Rd$ is analysed.
The twist is extended to 
quantum field theory, leading to the equivalences mentioned
above. Perturbative consequences are investigated.
Space-time and gauge symmetry are considered at the end.

The reader less familiar with the theory of
quantum groups is encouraged to
start by reading Section~\ref{sec:qft} and then return to
Sections~\ref{sec:twists} and \ref{sec:Rd} for the foundations.

\section{Foundations: Twists and Equivalence}
\label{sec:twists}

This section provides the necessary mathematical
foundations in the form needed for our treatment.
It is a review of known material except perhaps for the
two rather trivial Lemmas.
A useful standard reference for the general theory of quantum groups
and braided spaces is Majid's book \cite{Ma:book}.
We use the notations $\cop,\cou,\antip$ for coproduct,
counit and antipode of a Hopf algebra respectively. We use
Sweedler's notation $\cop a=a\i1\tens a\i2$ for coproducts and a
similar notation $v\mapsto v\i1\tens v\iu2$ for left coactions.
$\k$ denotes a general field.

We recall that a coquasitriangular structure $\Ru:H\tens H\to\k$ on
a Hopf algebra $H$ provides a braiding on its category of
comodules. More precisely, given any two comodules $V$ and $W$ there
is an intertwining map
$\psi:V\tens W \to W\tens V$ (the ``braiding'').
$\psi$ can be seen as a replacement of the concept of an ordinary
transposition which would be an intertwiner for representations of an
ordinary group. 
For left comodules, $\psi$ is given in terms of $\Ru$ as
\begin{equation}
 \psi(v\tens w)=\Ru(w\i1\tens v\i1)\, w\iu2\tens v\iu2 .
 \label{eq:psi}
\end{equation}
If $\Ru$ satisfies an extra condition, then for $H$-invariant elements
the braiding $\psi$ is just
the same as the flip map:
\begin{lem}
\label{lem:twobraid}
Let $H$ be a Hopf algebra with coquasitriangular structure $\Ru:H\tens
H\to\k$ satisfying the property $\Ru(\antip  a\i1\tens
a\i2)=\cou(a)$. Then for left comodules $V$ and $W$ and $v\tens w\in
V\tens W$ $H$-invariant we have
$\psi(v\tens w)=w\tens v$.
\end{lem}
\begin{proof}
$w\i1\tens v\i1\tens w\iu2\tens v\iu2
 = \antip v\i1\tens v\i2\tens w\tens v\iu3$ due to invariance.
Inserting this into
 (\ref{eq:psi}) gives the desired result.
\end{proof}
We note that this property extends to cyclic permutations of invariant
elements in multiple tensor products (just replace $V$ or $W$ by a
multiple tensor product). This fact will be of interest later.

We turn to the concept of a twist of a Hopf algebra.
Let $H$ be a Hopf algebra and $\chi: H\tens H\to\k$ be a unital
2-cocycle, i.e., a linear map that has a convolution inverse and
satisfies the properties
\begin{gather}
\chi(a\i1\tens b\i1)\,\chi(a\i2 b\i2\tens c)
=\chi(b\i1\tens c\i1)\,\chi(a\tens b\i2 c\i2) \label{eq:cocycle} \\
\text{and}\quad
\chi(a\tens 1)=\chi(1\tens a)=\cou(a) .
\label{eq:unital}
\end{gather}
This defines a twisted Hopf algebra $H'$ 
with the same coalgebra structure and unit as $H$. Its
product and antipode are given by
\begin{gather}
a\bullet b=\chi(a\i1\tens b\i1)\, a\i2 b\i2\,
 \chi^{-1}(a\i3\tens b\i3),
\label{eq:tprod} \\ 
\antip' a = U(a\i1)\antip a\i2 U^{-1}(a\i3)
\quad\text{with}\quad U(a)=\chi(a\i1\tens\antip a\i2) . \notag
\end{gather}
If $H$ carries a coquasitriangular structure $\Ru:H\tens H\to\k$, then
$H'$ carries an induced coquasitriangular structure $\Ru'$ given by
\begin{equation}
\Ru'(a\tens b)=\chi(b\i1\tens a\i1)\, \Ru(a\i2\tens b\i2)\,
 \chi^{-1}(a\i3\tens b\i3) .
 \label{eq:tcqtr}
\end{equation}

The main result we need in the following is that the twist that 
turns $H$ into $H'$ extends to the corresponding comodule categories
and establishes an equivalence. We formulate it here for left
comodules. It is ultimately due to Drinfeld. See \cite{Dri:quasi} for
a dual version in the setting of quasi-Hopf algebras.
\begin{thm}[Drinfeld]
\label{thm:catequiv}
Let $H$ be a Hopf algebra, $\chi:H\tens H\to\k$ a unital
2-cocycle, $H'$ the corresponding twisted Hopf algebra. There is an
equivalence of monoidal categories
$\funceq:\catmod{H}\to\catmod{H'}$. $\funceq$ leaves the
coaction
unchanged. The monoidal structure is provided by the natural
equivalence
\begin{align*}
\nateq:\funceq(V)\tens\funceq(W)&\to\funceq(V\tens W)\\
 v\tens w &\mapsto \chi(v\i1\tens w\i1)\, v\iu2\tens w\iu2
\end{align*}
for all $V,W\in\catmod{H}$.
If $H$ is coquasitriangular, the equivalence $\funceq$ becomes a
braided equivalence.
\end{thm}
Let us remark that an $H$-invariant element of a 2-fold
tensor product
remains the same under twist if $\chi$ satisfies an extra
property. This is the following Lemma.
\begin{lem}
\label{lem:twotwist}
In the context of Theorem~\ref{thm:catequiv} let $V$ and $W$ be
$H$-comodules and
$v\tens w\in V\tens W$ be $H$-invariant. Assume further that $\chi$
satisfies $\chi(a\i1\tens\antip a\i2)=\cou(a)$. Then
$\nateq^{-1}(v\tens w)=v\tens w$.
\end{lem}
\begin{proof}
First observe that the mentioned property of $\chi$ is automatically
satisfied by $\chi^{-1}$ as well. Then use invariance in the form
$v\i1\tens w\i1\tens v\iu2\tens w\iu2 
 = v\i1\tens \antip v\i2\tens v\iu3\tens w$ and apply $\nateq^{-1}$.
\end{proof}
The equivalence $\nateq$ extends to multiple tensor products by
associativity. We denote the extension to an $n$-fold tensor product
by $\nateq_n$. The induced transformation of a morphism $\alpha$
sending an $n$-fold to an $m$-fold tensor is
$\nateq_m^{-1}\circ\alpha\circ\nateq_n$.
In particular, we can apply this to the product map of an
algebra (see \cite{Ma:book} for similar examples).
\begin{ex}
\label{ex:algtwist}
Let $H$ be a Hopf algebra, $A$ a left $H$-comodule algebra and $\chi$
a unital 2-cocycle over $H$. Then $A'$ built on $A$ with the new
multiplication
\[
a\star b=\chi(a\i1\tens b\i1)\, a\iu2 b\iu2
\]
is an $H'$-comodule algebra. Note that the associativity of the
product is ensured by the cocycle condition (\ref{eq:cocycle}).
\end{ex}

It is well known that for a Lie group $G$, a twist of its Hopf algebra
of functions provides a (strict) deformation quantisation.
More interestingly in our context, a twist also provides a deformation
quantisation on
any manifold $M$ that $G$ acts on. In particular, taking $M=G$
leads to a different (non-strict) deformation
quantisation. Recall that a deformation quantisation of a manifold
$M$ is an associative linear map
$\star:\falg(M)\tens\falg(M)\to\falg(M)[[\hbar]]$ which satisfies
$f\star g= f g+\O(\hbar)$ and $f\star g- g\star f= \hbar
\{f,g\}+\O(\hbar^2)$ where $\{\cdot,\cdot\}$ is a Poisson bracket on
$M$. One usually also requires that the $\star$-product is defined
for all orders in $\hbar$ by bidifferential operators (see e.g.\
\cite{BFFLS:defquant}). 
The following Proposition (in dual form) is due to Drinfeld
\cite{Dri:constybe}.
\begin{prop}[Drinfeld]
\label{prop:defquant}
Let $G$ be a Lie group acting on a manifold $M$. Denote by
$H=\falg(G)$ the (topological) Hopf algebra of functions on $G$ and by
$A=\falg(M)$ the
$H$-comodule algebra of functions on $M$. Then a unital 2-cocycle 
$\chi:H\tens H\to\C[[\hbar]]$ so that $\chi(f \tens h)=\cou(f)\cou(h)
+\O(\hbar)$ defines a deformation quantisation on $A$.
\end{prop}

Finally, let us mention that for commutative Hopf algebras $C$ and $H$
with $C$ a left $H$-comodule algebra and coalgebra there is a commutative
semidirect product Hopf algebra $C\rtimes H$. It is freely generated
by $C$ and $H$ as a commutative algebra. Its coproduct on elements of
$H$ is the
given one, while the coproduct on elements of $C$ is modified to
\begin{equation}
 \cop_{\rtimes} c = c\i1 c\i2\ib1 \tens c\i2\ibu2 .
\label{eq:sdprod}
\end{equation}
Here brackets denote
the coaction to distinguish
it from the coproduct.
This is the straightforward equivalent to a semidirect product of
groups in quantum group language. For the general theory of crossed
products of Hopf algebras see \cite{Ma:book}.
\section{Noncommutative $\Rd$ as a Twist}
\label{sec:Rd}

Part of the discussion in this section reproduces
\cite{Wat:ncstringhopf}. In particular, the 2-cocycle
(\ref{eq:ncdtwist}) was
found there, and it was shown to give rise to the deformed
product (\ref{eq:ncdef}). However, the full representation
theoretic picture essential to our treatment of quantum
field theory was lacking.
We provide it here.

We work over the complex numbers from now on.
Although we use the purely algebraic language for convenience,
Hopf algebras are to be
understood in a topological sense in the following.
Tensor products are appropriate completions.
One could use the setting of Hopf $C^*$-algebras for example
\cite{VaVD:hopfcstar}.
However, our discussion is independent of the functional
analytic details and so we leave them out.
When referring to function algebras one should have in mind a class
compatible with the functional analytic setting chosen.

Consider $\Rd$ as the group of translations of $d$-dimensional
Euclidean space. In the language of quantum groups, the corresponding
object is the Hopf algebra $\CR$ of functions on $\Rd$.
We can view this as (a certain completion of) the unital commutative
algebra generated by the coordinate functions $\{x^1,\dots,x^d\}$.
The product is $(f\cdot g)(x)=f(x)\cdot g(x)$,
the unit $1(x)=1$, the counit $\cou(f)=f(0)$, and the antipode $(\antip
f)(x)=f(-x)$.
Identifying the (completed) tensor product $\CR\tens\CR$ as the
functions on the cartesian product $\falg(\Rd\times\Rd)$ the
coproduct encodes the group law of translation via 
$\cop(f)(x,y)=f(x+y)$.
We can formally write this as a Taylor expansion
\[
 \cop f = \exp\left(x^\mu\tens \pdiff{x^\mu}\right) (1\tens f)
 = \exp\left(\pdiff{x^\mu}\tens x^\mu\right) (f\tens 1) .
\]
We have the usual $*$-structure $(x^\mu)^*=x^\mu$ making $\CR$ into a
Hopf $*$-algebra.
$\CR$ is naturally equipped with the trivial coquasitriangular
structure $\Ru=\cou\tens\cou$.

Taking the dual point of view, we consider the Lie algebra
of translation generators with basis $\{p_1,\dots,p_d\}$. We denote
its 
universal envelope by $\UR$. Expressing elements of $\UR$ as functions
in the $p_\mu$, we obtain the same Hopf algebra structure as for
$\CR$. We define the dual pairing by
\[
\langle f(p_\mu),g\rangle
 = f\left(i\pdiff{x^\mu}\right) g(x)\Big|_{x=0} .
\]
The corresponding (left) action of $\UR$ on $\CR$ that leaves this
pairing invariant is given by
\[
(p_\mu\act g)(x)=-i\pdiff{x^\mu} g(x) .
\]
Viewing $\UR$ as momentum space, we have the usual translation
covariant Fourier transform
$\hat{}:\CR\to\UR$ and its inverse given by
\begin{equation}
\hat{f}(p)=\int \frac{\xd^d\!x}{(2\pi)^{d/2}} f(x)
 e^{-ip_\mu x^\mu}\quad \text{and} \quad
f(x)=\int \frac{\xd^d\!p}{(2\pi)^{d/2}} \hat{f}(p)
 e^{ip_\mu x^\mu} .
\label{eq:ft}
\end{equation}

Now, let $\theta$ be a real valued antisymmetric $d\times d$
matrix. Consider the map $\chi_\theta:\CR\tens\CR\to\C$ given by
\begin{equation}
 \chi_\theta(f\tens g)=(\cou\tens\cou)\circ
  \exp\left(\frac{i}{2}\,\theta^{\mu\nu}
  \pdiff{x^\mu}\tens\pdiff{x^\nu}\right) (f\tens g) .
 \label{eq:nctwist}
\end{equation}
One easily verifies (check (\ref{eq:cocycle}) and (\ref{eq:unital}))
that this defines a unital
2-cocycle on $\CR$ with inverse
$\chi_\theta^{-1}=\chi_{-\theta}=
\chi_\theta\circ\tau$ ($\tau$ the flip map).
Thus, according to section~\ref{sec:twists} it gives rise to a twisted
Hopf algebra $\CRt$.
However, the twisted product is the same as the original product,
i.e., $\CR$ and $\CRt$ are identical as Hopf algebras.
In other words -- the group of translations remains unchanged.
In fact, it is easy to see from (\ref{eq:tprod}) that this must be so
for any twist 
on a cocommutative Hopf algebra. The coquasitriangular structure does
change on the other hand, and we obtain
\[
 \Ru_\theta(f\tens g)
  = (\cou\tens\cou)\circ\exp\left(-i\,\theta^{\mu\nu}
  \pdiff{x^\mu}\tens\pdiff{x^\nu}\right) (f\tens g) .
\]
according to (\ref{eq:tcqtr}).
In particular, this means that the category of
comodules of $\CRt$ is equipped with a braiding $\psi_\theta$ that is
not the flip map. Using (\ref{eq:psi}) we obtain
\begin{equation}
 \psi_\theta(f\tens g)=\exp\left(-i\,\theta^{\mu\nu}
  \pdiff{x^\mu}\tens\pdiff{x^\nu}\right) (g\tens f) .
\label{eq:ncbraid}
\end{equation}
In more conventional language this means that the
representations of the translation group aquire non-trivial
statistics.
Note that
$\Ru_\theta^{-1}=\Ru_\theta\circ\tau$ (with $\tau$ the flip map), i.e.,
$\Ru_\theta$ is cotriangular (as it must be, being obtained by twisting
from a trivial $\Ru$). Consequently, the braiding is symmetric, i.e.,
$\psi_\theta^2=\id$.

By duality we can equivalently express this twist as an invertible
element
$\Phi_\theta\in\UR\tens\UR$ obeying the dual axioms of
(\ref{eq:cocycle}) and (\ref{eq:unital}).
We get
\begin{equation}
 \Phi_\theta=\exp\left(-\frac{i}{2}\,\theta^{\mu\nu}
  p_\mu\tens p_\nu\right) .
\label{eq:ncdtwist}
\end{equation}
This is (3.10) in \cite{Wat:ncstringhopf}.
As in the above discussion the twisted $\URt$
is the same as 
$\UR$ as a Hopf algebra, but the quasitriangular structure becomes
nontrivial.

Now consider $d$-dimensional Euclidean space with an action of the
translation group (from the left say). In quantum group language
this means that we take a second copy $\CRc$
of $\CR$ as a left $\CR$-comodule
algebra. In contrast to the quantum group $\CR$ its algebra structure
\emph{is} changed under the twist. 
This is the situation of Example~\ref{ex:algtwist}.
Furthermore, we know from Proposition~\ref{prop:defquant} that the new
product on
the twisted $\CRc$ which we denote by $\CRct$ is a
deformation quantisation.
We find
\begin{equation}
 (f\star g)(x)=\exp\left(\frac{i}{2}\,\theta^{\mu\nu}
  \pdiff{\xi^\mu}\pdiff{\eta^\nu}\right)
  f(x+\xi)g(x+\eta)\Big|_{\xi=\eta=0} ,
\label{eq:ncdef}
\end{equation}
which is known as a Moyal $\star$-product \cite{Moy:quantstat}.
Note that the inherited $*$-structure is compatible
with the new algebra structure making $\CRct$ into a $*$-algebra.

According to Theorem~\ref{thm:catequiv}, the category
of $\CR$-comodules and the category
of $\CRt$-comodules are equivalent. While objects remain the same
under twisting the tensor product does not. In particular, this means
that while for $f\in\CRc$ the corresponding $f_\theta\in\CRct$ is just
the same function this is not so for functions of several variables.
In our context a function of $n$ variables is an element of
$\widetilde{\falg}(\Rd\times\cdots\times\Rd)$ which we write as the
tensor product
$\CRc\tens\cdots\tens\CRc$. 
This is transformed to the tensor product
$\CRct\tens\cdots\tens\CRct$ via the functor
$\nateq^{-1}_n$.
Explicitly, we obtain
\begin{equation}
 f_\theta(x_1,\dots,x_n)=\exp\left(-\frac{i}{2}\,\sum_{l<m}\theta^{\mu\nu}
  \pdiff{x^\mu_l}\pdiff{x^\nu_m}\right) f(x_1,\dots,x_n) .
\label{eq:equivp}
\end{equation}
Due to duality (left) $\CR$-comodules are really the same thing as
(left) $\UR$-modules. In particular, viewing momentum space as a
left $\UR$-module (coalgebra) denoted by $\URc$, it lives in the same
category as $\CRc$ and we denote its twisted analogue by $\URct$. The
momentum space version of equation (\ref{eq:equivp}) reads
\begin{equation}
 f_\theta(p^1,\dots,p^n)=\exp\left(\frac{i}{2}\,\sum_{l<m}\theta^{\mu\nu}
  p_\mu^l p_\nu^m\right) f(p^1,\dots,p^n) .
\label{eq:equivm}
\end{equation}
The transformation of morphisms (i.e.\ intertwiners) by $\funceq$ is
non-trivial only if they transform tensor products to tensor
products. In particular, this means that integration and Fourier
transform (\ref{eq:ft}) are preserved by the twist. Note that even the
Fourier transform in
several variables survives the twist unchanged, since it factors
into Fourier transforms in each variable.

\subsection{A Remark on the Noncommutative Torus}
\label{sec:torus}

All constructions we have made for noncommutative $\Rd$ apply
equally to the noncommutative torus. We simply
restrict to periodic functions. To be more specific, let $\Td$
denote the group $U(1)^d$ of translations on the $d$-dimensional
torus of unit radius which we also denote by $\Td$. The Hopf algebra
of functions
$\CT$ on $\Td$ has a basis of Fourier modes $\{u_k\}$ for
$k\in\Z^d$. We can identify $u_k$ as a periodic function in $\CR$ via
$u_k(x)= \exp(i\,k_\mu x^\mu)$. 
For completeness we provide the relevant formulas explicitly:
Product and coproduct are given
by $u_k u_l= u_{k+l}$ and $\cop u_k=u_k\tens u_k$.
The counit is $\cou(u_k)=1$. Antipode and $*$-structure are $\antip
u_k=u_k^*= u_{-k}$.
The twist (\ref{eq:nctwist}) takes the form
\[
 \chi_\theta(u_k\tens u_l)=\exp\left(-\frac{i}{2}\,\theta^{\mu\nu}
  k_\mu l_\nu\right)
\]
and the twisted comodule algebra $\CTct$ satisfies the product
rule
\[
 u_k \star u_l = \exp(-i\,\theta^{\mu\nu} k_\mu l_\nu)\,
  u_l \star u_k .
\]

\section{Quantum Field Theory on Noncommutative $\Rd$}
\label{sec:qft}

Let us examine the noncommutative $\Rd$ with a
view towards taking it as the space-time of a quantum field theory.
Recall that 
the coordinate functions $x^1,\dots,x^d$ 
of noncommutative $\Rd$ obey commutation relations of the form
\begin{equation}
 [x^\mu,x^\nu]=i\,\theta^{\mu\nu}
\label{eq:ncRd}
\end{equation}
for $\theta$ a real valued antisymmetric $d\times d$ matrix.
More precisely, noncommutative $\Rd$ is a deformation quantisation of the
algebra of functions on ordinary $\Rd$ satisfying (\ref{eq:ncdef}).

Apart from space-time itself, its group of isometries plays a
fundamental role in quantum field theory.
After all, fields and particles are representations of this group (or
its universal cover) and it leaves a quantum field theory as a whole
(i.e., its $n$-point functions) invariant.
What is this group for noncommutative $\Rd$?
For general $\theta$,
the commutation relations (\ref{eq:ncRd}) are clearly not invariant
under rotations or boosts. However, they are invariant under ordinary
translations $x^\mu \mapsto x^\mu+a^\mu$.
Thus, it appears natural to let the translations play the role of
isometries of noncommutative $\Rd$.
This is an important ingredient for the following discussion. We later
come back to
the question of a possible larger group of symmetries.

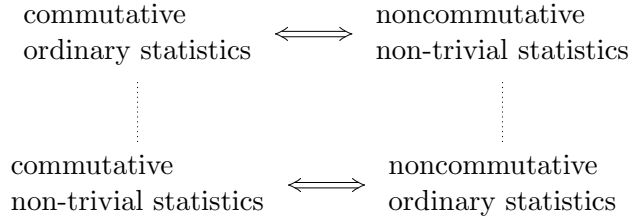
\begin{figure}
\begin{center}
$\sxy{\xymatrix{
{\begin{tabular}{l}
 commutative\\ ordinary statistics
 \end{tabular}}\ar@{<=>}[r]\ar@{.}[d] &
{\begin{tabular}{l}
 noncommutative\\ non-trivial statistics
 \end{tabular}}\ar@{.}[d] \\
{\begin{tabular}{l}
 commutative\\ non-trivial statistics
 \end{tabular}}\ar@{<=>}[r] &
{\begin{tabular}{l}
 noncommutative\\ ordinary statistics
 \end{tabular}}}}$

\caption{Relations between quantum field theories on $\Rd$.
 The arrows indicate equivalences while the dotted lines indicate
 equality of planar Feynman diagrams.}
\label{fig:duality}
\end{center}
\end{figure}
It was shown in Section~\ref{sec:Rd} how ordinary $\Rd$ is turned into
noncommutative $\Rd$ by a process of twisting.
This is induced by a 2-cocycle $\chit$ on the quantum group $\CR$ of
translations. (``Cocycle'' here has the meaning dual to that of
ordinary group cohomology.)
At the same time $\CR$ is turned into the
quantum group $\CRt$. While this still corresponds to the ordinary
group of translations, it is different from $\CR$ as a quantum
group. The difference is encoded in the coquasitriangular structure
$\Rut$ which is now non-trivial. It equips noncommutative $\Rd$ with
a non-trivial statistics encoded in the braiding $\psit$. This
twist-transformation is
represented in Figure~\ref{fig:duality} by the upper arrow. It goes
both ways since we can undo the twist by using the inverse 2-cocycle
$\chimt$.

What about noncommutative $\Rd$ with \emph{ordinary}
statistics?
After all, this is the space which has been of interest in the
literature. Untwisting this space yields the commutative $\Rd$ as
before. However, as before, twisting also exchanges ordinary with
braided statistics. Only this time the other way round: We obtain
commutative $\Rd$ equipped with braided statistics. This is represented
by the lower arrow in Figure~\ref{fig:duality}.
In the language of Section~\ref{sec:Rd}, we consider $\CRct$
(noncommutative $\Rd$) as
a comodule of $\CR$ (the translation group with ordinary statistics)
and apply the twist with the inverse 2-cocycle
$\chimt$. We get $\CRc$ (ordinary $\Rd$) but as a comodule of $\CRmt$
(the translation group with braided statistics).
The braiding this time is given by $\psimt$ since we have
used the inverse twist. Note that the braiding is in both cases
symmetric, i.e., $\psi^2$ is the identity.

We show in the following how the twist equivalence between the
respective spaces gives rise to an equivalence of quantum field
theories on those spaces.
In order to do that we need to express (perturbative)
quantum field theory in a purely algebraic language. Also, we need to
be able to deal with quantum field theory on spaces carrying a braided
statistics. Both is accomplished by braided quantum field theory
\cite{Oe:bQFT}. This
is briefly reviewed in Section~\ref{sec:symqft} and specialised for the
case of a symmetric braiding.
We can then go on to show the equivalences in
Section~\ref{sec:equivqft} and look in more detail at the perturbative
consequences in Section~\ref{sec:pert}.
Finally, in Section~\ref{sec:symmetry}, we turn to the question of
what happens with additional symmetries under twist.

Note that while the whole discussion is solely in terms of $\Rd$ for
convenience, everything applies equally to the
torus. This follows from the remarks in Section~\ref{sec:torus}.
The only exception are the extra space-time symmetries considered in
Section~\ref{sec:stsym}.

\subsection{Symmetric Braided Quantum Field Theory}
\label{sec:symqft}
\begin{figure}
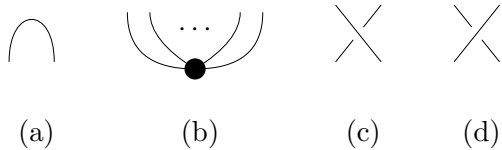

\begin{center}
\begin{tabular}{ccccccc}
\input{fig_arch}
& &
\input{fig_vertex}
& &
\input{fig_over}
& &
\input{fig_under}
\\ \\
(a) & & (b) & & (c) & & (d)
\end{tabular}
\caption{Free propagator (a) and vertex (b). Over-crossing (c) and
under-crossing (d).}
\label{fig:elements}
\end{center}
\end{figure}
\begin{figure}
\begin{center}
\input{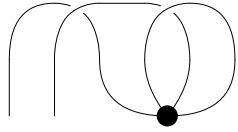}
\caption{One of the tadpole diagrams in braided quantum field theory.}
\label{fig:btadpole}
\end{center}
\end{figure}
Braided quantum field theory employs the same path integral
formulation and perturbation expansion as ordinary quantum field
theory. However, one has to be much more careful since
ordering is relevant even inside the path integral due to the braided
statistics of the underlying space.
As a consequence, one needs to impose extra restrictions to the way
Feynman diagrams are drawn.
It was shown in \cite{Oe:bQFT} how this is accomplished by
using an adapted version of the diagrammatic language for
braided categories. Let us briefly recall the rules:
Diagrams are drawn such that all external legs end on the bottom
line. Vertices are drawn with all legs pointing upwards (see
Figure~\ref{fig:elements}.b). Free 
propagators are arches (Figure~\ref{fig:elements}.a) connecting
vertices and/or external legs. 
Over- and under-crossings of lines (Figure~\ref{fig:elements}.c\&d)
are distinct.
Diagrams are evaluated from top to bottom.
Horizontally parallel strands correspond to tensor
products with each strand representing a field. Denoting the space
of fields by $X$, a free propagator is an element of $X\tens X$.
A vertex with $n$ legs is a map $X\tens\cdots\tens X\to\k$ with the
tensor product being $n$-fold.
Over- and under-crossings correspond to the braiding
map $\psi:X\tens X\to X\tens X$ and its inverse, encoding the
non-trivial statistics.
The diagram as a whole is an element of
$X\tens\cdots\tens X$ with as many factors as external lines.
Notice that propagators, vertices, and diagrams are by construction
invariant under the given (quantum) group of space-time symmetries.
As an example, Figure~\ref{fig:btadpole} shows a diagram contributing
to the 2-point function at 1-loop order in $\phi^4$-theory.
It corresponds to a
tadpole diagram (Figure~\ref{fig:tadpole}.b) in ordinary quantum field
theory.

In the case of a symmetric braiding, over- and under-crossings
become the same and
the situation is simplified considerably.
The complication in the general braided case really is that a
permutation of components in a tensor product depends on what
transpositions (using $\psi$ or $\psi^{-1}$) have occured and in which
order. This defines a representation of the braid group. In the
symmetric case however, any sequence of transpositions (using
$\psi=\psi^{-1}$) leading to a given permutation is equivalent. Thus,
we have a
representation of the symmetric group. For the Feynman diagrams this
means that we can return to the usual freedom of ordinary quantum
field theory in drawing them.
Any way of rearranging an ordinary Feynman graph so that it conforms
with the stricter rules of general braided Feynman graphs is
equivalent due to the symmetry of the braiding.
This supposes that the free propagator is invariant under
the braiding $\psi$. (This is satisfied in the concrete cases
considered.)

An analogous simplification occurs with respect to the perturbation
expansion. In ordinary quantum field theory the combinatorics of which
diagrams are generated is encoded in Wick's theorem. In braided quantum
field theory the corresponding role is played by the braided
generalisation of Wick's
theorem \cite{Oe:bQFT}. In the symmetric case, this again
reduces to the ordinary Wick's theorem.

While in the general braided case a diagram is evaluated strictly from
top to bottom, this can be relaxed to the ordinary way of evaluating
a diagram for the symmetric case. With one crucial
exception: Every crossing of lines in a diagram is associated with the
braiding $\psi$. If $\psi$ is not just the flip map, we obtain an
extra Feynman rule for crossings. (For the commutative $\Rd$ with
braided statistics this is e.g.\ the rule depicted in
Figure~\ref{fig:extra}.)

\subsection{The Equivalences for Quantum Field Theory}
\label{sec:equivqft}

With the machinery of symmetric braided quantum field theory in place
we can handle quantum field theory on any of the versions
of $\Rd$ represented in Figure~\ref{fig:duality}.
Recall that the arrows in this figure represent twist transformations
between the respective spaces. Now, by Theorem~\ref{thm:catequiv} the
twist induces an equivalence between the whole categories
of translation covariant objects and maps in which those spaces live.
But, as demonstrated in the previous section, the whole perturbation
expansion takes place in this category, including Feynman diagrams and
$n$-point functions.
This is made explicit by using braided Feynman
diagrams and associating the space of fields, tensor products and
intertwining maps (vertices, the braiding etc.) with elements of those
diagrams.
Consequently, quantum field theories on spaces related by twist are
equivalent. In particular, the arrows in Figure~\ref{fig:duality}
stand for such equivalences.
For an $n$-point function, a Feynman diagram or a vertex the relation
between the commutative quantity $G$ and the noncommutative quantity
$G_{NC}$ is in both cases
given in momentum space by
\begin{equation}
 G_{NC}(p^1,\dots,p^n)=\exp\left(\frac{i}{2}\,\sum_{l<m}\theta^{\mu\nu}
  p_\mu^l p_\nu^m\right) G(p^1,\dots,p^n)
\label{eq:equivqft}
\end{equation}
which is just formula (\ref{eq:equivm}) from Section~\ref{sec:Rd}. The
corresponding position space version is (\ref{eq:equivp}).

We would like to stress that our treatment applies to fields in any
representation of the translation group
and thus to quantum field theory in general.
For scalars the space of fields is simply $\CRct$ itself. Any other
field lives
in a bundle associated with the frame bundle (or its universal cover --
the spin bundle) which in particular carries an action of the
translation group. Choosing a trivialisation ``along translation''
allows to write the space of sections of the bundle as $V\tens \CRc$
with translations acting trivially on $V$. Thus, under twist we obtain
$V\tens \CRct$ with the $V$-component not being affected at all by the
twist. In other words: Extra indices like spinor or tensor indices
just show the ordinary behaviour and can be considered completely
separate from the noncommutativity going on in space-time.
This also applies to the anticommutation of fermions.

Let us make an extra remark about gauge theories. 
For a gauge bundle there is no canonical action of the translation
group. Choosing such an action
is the same thing as choosing a trivialisation,
i.e., a ``preferred gauge''.
Given such a choice we can
treat gauge theory with the above methods.
This supposes that we have integrated out the gauge degrees of freedom
in the path integral in the usual way,
say by the Faddeev-Popov method.

Rigourously speaking, our treatment so far has assumed that quantities
encountered in the calculation of Feynman diagrams are finite. Then
the transformation (\ref{eq:equivqft}) between quantum
field theories connected by arrows in Figure~\ref{fig:duality} is
straightforward.
In order to establish the
equivalence not only for finite but also for renormalisable quantum
field theories,
we need to extend the twisting equivalence to the 
regularisation process involved in the renormalisation.
The only condition for the twist transformation to work in this
context is that we remain in the translation covariant category, i.e.,
that the regularisation preserves covariance under translations.
This is easily accomplished. For example, a simple momentum-cutoff
regularisation would do, or a
Pauli-Villars regularisation. (Note however, that the popular
dimensional regularisation can not be used here.)
Using such a regularisation, the
twisting equivalence holds at every step of the
renormalisation procedure, in
particular for the renormalised quantities at the end.

As a further remark, the equivalences should also hold
non-perturbatively, since the $n$-point functions (perturbative or
not) naturally live in the respective categories. However, for lack of
a general non-perturbative method, we can obviously
not demonstrate this explicitly.
Turning the argument round, one could say that for a well defined
theory on one side the transformation (\ref{eq:equivqft})
\emph{defines} the respective equivalent theory.

\subsection{Perturbative Consequences}
\label{sec:pert}
\begin{figure}
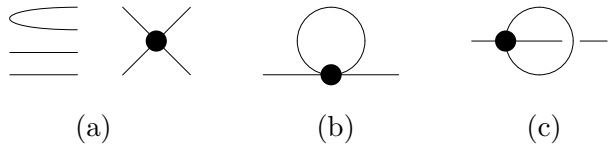

\begin{center}
\begin{tabular}{ccccc}
\input{fig_built}
& &
\input{fig_tadpole}
& &
\input{fig_ctadpole}
\\ \\
(a) & & (b) & & (c)
\end{tabular}
\caption{Building blocks for the diagrams of the first order
contribution to the 2-point function in $\phi^4$-theory (a).
Resulting tadpole diagram (b). In the cyclic case diagram (c) is
non-equivalent.}
\label{fig:tadpole}
\end{center}
\end{figure}
\begin{figure}
\begin{center}
\input{fig_crossingrule}
\caption{Extra Feynman rule for crossings with braided statistics.
 The arrows indicate the direction of the momenta $p$ and $q$.}
\label{fig:extra}
\end{center}
\end{figure}
Let us explore the consequences of the equivalences in terms of
perturbation theory. We first discuss the issue of vertex symmetry.
It has been observed that vertices which are totally symmetric under
an exchange of legs retain only a cyclic symmetry on noncommutative
$\Rd$ (with ordinary statistics). Following the upper arrow in
Figure~\ref{fig:duality} from left to right
we retain total symmetry. For a
transposition this takes the form
\begin{multline*}
 G_{NC}(p^1,\dots,p^i,p^{i+1},\dots,p^n) \\
 = \exp\left(-i\,\theta^{\mu\nu} p_\mu^i p_\nu^{i+1}\right)
  G_{NC}(p^1,\dots,p^{i+1},p^i,\dots,p^n) .
\end{multline*}
However, ``stripping off'' the non-trivial braiding, i.e., considering
ordinary transpositions by flip, leaves only a cyclic symmetry.
Following the lower arrow to the left, we have the opposite
situation. Vertices are now ordinarily totally symmetric, but we have a
non-trivial braiding with respect to which they are only cyclic
symmetric.

The deeper reason for the retention of cyclic symmetry is
a property of the coquasitriangular structure $\Rut$ defining the
braiding. As a consequence of this property, cyclic symmetry with
respect to ordinary and the braided statistics is the same for
translation invariant objects like vertices. This is
Lemma~\ref{lem:twobraid}.
For perturbation theory the use of vertices that are only cyclic
symmetric means that diagrams which would be the same for total
symmetry may now differ.
Consider for illustration the 2-point function in $\phi^4$-theory at 1-loop
order. Assembling the building blocks (Figure~\ref{fig:tadpole}.a) in
all possible ways (noting 
that the legs of the propagator are to be considered identical)
results in 8 times diagram (b) plus 4 times diagram (c),
given only cyclic symmetry of the vertex (see
Figure~\ref{fig:tadpole}). 
A total symmetry would imply that both
diagrams are equal, leading to the usual factor of 12. 

Now, recall from Section~\ref{sec:symqft} that a non-trivial braided
statistics leads to the appearance of an extra
Feynman rule. The braiding map $\psi$ instead of the trivial exchange
is now associated with each crossing.
In fact, this is the only effect of the (symmetric) braiding in
perturbation theory. It follows immediately that planar Feynman
diagrams are identical in theories that differ only by their
(symmetric) braided statistics. This is indicated by the dotted lines
in Figure~\ref{fig:duality}.

Filk's result \cite{Fil:qspace} for planar
diagrams is an immediate consequence:
We evaluate a planar diagram
in the commutative setting
and follow the lower arrow in Figure~\ref{fig:duality} to the right.
Diagrams are simply related by the equivalence formula
(\ref{eq:equivqft}).
For non-planar diagrams we also use the commutative setting.
We only
have to take into account the crossing factors
from the non-trivial statistics. They are given by the extra Feynman
rule in Figure~\ref{fig:extra}. This is the momentum space version of
formula (\ref{eq:ncbraid}) with opposite sign for $\theta$.
If we aggregate the factors
for a given diagram
by encoding all the crossings into an intersection matrix, we obtain an
overall factor 
\begin{equation}
\exp\left(i\,\sum_{k>l} I_{k l} \theta^{\mu\nu}
 p_\mu^k p_\nu^l\right) .
\label{eq:nonplan}
\end{equation}
Here, the indices $k,l$ run over all lines of the diagram and
$I_{k l}$ counts the oriented number of intersections between lines $k$
and $l$. Then again, relation (\ref{eq:equivqft}) leads to the
noncommutative theory.
This is Filk's result for non-planar diagrams \cite{Fil:qspace}.
Note that it was already observed in \cite{MVS:ncpert} that
(\ref{eq:nonplan})
can be obtained by assigning phase factors to crossings.

As a further remark, it has been observed that quadratic terms in
the Lagrangian are not modified in the noncommutative setting.
This follows from a property of the twisting cocycle.
Any invariant object with 2 components (like a 2-leg vertex, a free
propagator etc.) remains unchanged by the
twist. This is Lemma~\ref{lem:twotwist}.

\subsection{Additional Symmetries}
\label{sec:symmetry}

In this final section we consider the effect of twisting on additional
symmetries.
We follow the upper arrow in Figure~\ref{fig:duality} from
left to right.

\subsubsection{Space-Time Symmetry}
\label{sec:stsym}

As mentioned before, the commutation relations
(\ref{eq:ncRd}) are not
invariant under rotations. However, ordinary Euclidean space is, and
since the noncommutative version is a twist of the commutative one,
there should be an analogue of those symmetries. This is indeed the
case. Consider the group of (orientation preserving) rotations $SO(d)$
in $d$ 
dimensions. In quantum group language we consider the algebra of
functions $\falg(SO(d))$ generated by the matrix elements $t^\mu_\nu$
of the fundamental representation. We have relations $t^\mu_\rho
t^\nu_\rho=\delta^\mu_\nu=t^\rho_\mu t^\rho_\nu$ (summation over
$\rho$ implied) and $\det(t)=1$, coproduct
$\cop t^\mu_\nu=t^\mu_\rho \tens t^\rho_\nu$, counit
$\cou(t^\mu_\nu)=\delta^\mu_\nu$, and antipode $\antip
t^\mu_\nu=t^\nu_\mu$. We have a $*$-structure given by
$(t^\mu_\nu)^*=t^\nu_\mu$.

We extend the translation group $\Rd$ to the full
group $E\defeq\Rd \rtimes SO(d)$ of (orientation preserving)
isometries of Euclidean space. I.e.\ we consider the Hopf algebra
$\falg(E)=\falg(\Rd\rtimes SO(d))\cong \CR\rtimes\falg(SO(d))$. 
The rotations
(co)act on the translations from the left by $x^\mu\mapsto
t^\mu_\nu\tens x^\nu$. The resulting semidirect product Hopf algebra
is generated by $x^\mu$ and $t^\mu_\nu$ with the given relations. The
coproduct of $t^\mu_\nu$ remains the same but for $x^\mu$ we now
obtain $\cop x^\mu= x^\mu\tens 1 + t^\mu_\nu\tens x^\nu$. (Use
(\ref{eq:sdprod}).) This also determines the left coaction on $\CRc$.

The cocycle $\chi_\theta$ on $\CR$ extends trivially to a cocycle on
the larger quantum group $\falg(E)$, i.e., we let
$\chi_\theta$ just be the counit on the generators of
$\falg(SO(d))$. The twist \emph{does} change the 
algebra structure now. This was to be expected since we have already
seen that ordinary rotation invariance is lost. What do we have
instead? Using (\ref{eq:tprod}) and (\ref{eq:nctwist}) we find that
the relations for the $x^\mu$ become
\[
 x^\mu\bullet x^\nu - x^\nu\bullet x^\mu
 = i\,\theta^{\mu\nu}
   - i\,\theta^{\rho \sigma} t^\mu_\rho\bullet t^\nu_\sigma ,
\]
while the $t^\mu_\nu$ still commute with the other generators.
Thus, the twisted space-time symmetries $\falg_\theta(E)$
form a genuine quantum group (noncommutative Hopf algebra), no longer
corresponding to any ordinary group.

When dealing with the translation group alone, we were able to 
remove the non-trivial coquasitriangular structure responsible for the
braided statistics and replace it by a trivial one (follow the
dotted line on the right in Figure~\ref{fig:duality}
downwards). However, this is
no longer possible for the whole Euclidean motion group. A
genuine quantum group as the one obtained here does not admit a
trivial coquasitriangular structure.
Thus, removing the braided statistics really breaks the symmetry for the
quantum field theory.
(Note that the argument applies to Minkowski space and the Poincar\'e
group in the identical way.)

\subsubsection{Gauge Symmetry}

Let us consider a gauge theory with gauge group $G$. The
gauge transformations are the maps $\Rd\to G$. We denote the group of
such maps by $\Gamma=\{\Rd\to G\}$. The symmetry group generated by
translations and gauge transformations is the semidirect product
$\Omega\defeq\Gamma\rtimes\Rd$, where we have chosen an action of the
translation group on the gauge bundle.
(Note that the inclusion of further space-time symmetries does not
modify the argument.)
While the group $\Omega$ ``forgets'' about the
trivialisation of the gauge bundle corresponding to the chosen action,
we do need the trivialisation to
extend the twisting cocycle from $\Rd$ to $\Omega$. This is in
accordance with our remark on gauge theories above.
In quantum group language we have the semidirect
product of Hopf algebras $\falg(\Omega)=\falg(\Gamma\rtimes\Rd)\cong
\falg(\Gamma)\rtimes\CR$. The cocycle $\chi_\theta$ extends trivially
from $\CR$ to $\falg(\Gamma)$.
Applying the twist (\ref{eq:tprod}) with (\ref{eq:sdprod})
results in a noncommutative product
\begin{gather*}
 f\bullet\gamma = \chi_\theta(f\i1\tens\gamma\i1)\,
  f\i2\gamma\iu2 ,  \\
 \gamma\bullet\omega = \chi_\theta(\gamma\i1\tens\omega\i1)\,
  \gamma\iu2\omega\iu2 ,
\end{gather*}
while $f\bullet g= f g$ for $f,g\in\CR$ and
$\gamma,\omega\in\falg(\Gamma)$. Thus, the group of gauge
transformation does not survive the twist as an ordinary group. As for
the case of rotations we find that we obtain a genuine quantum group.
Again, the removal of the braided statistics would break the symmetry.
Note that this does not exclude the possibility of different gauge
symmetries. See \cite{CoRi:ymnctorus} and more recently
\cite{SeWi:stringncg} for discussions of gauge theory on
noncommutative $\Rd$.
\section*{Acknowledgements}

I would like to thank Shahn Majid for valuable discussions during the
preparation of this paper. 
I would also like to acknowledge financial support by
the German Academic Exchange Service (DAAD) and
the Engineering and Physical Sciences Research Council (EPSRC).

\bibliographystyle{myphys}
\bibliography{references}
\end{document}